\documentclass{elsart1p}
\usepackage{epsfig}
\usepackage{graphicx}
\usepackage[figuresright]{rotating}
\usepackage{amssymb}
\begin{document}
\begin{frontmatter}
%
%
%
\title{Physics Beyond the Standard Model}
%
%
\author{John Ellis}
\address{Theory Division, Physics Department, CERN, CH 1211 Geneva 23, Switzerland}
\begin{abstract}
The Standard Model is in good shape, apart possibly from $g_\mu - 2$
and some niggling doubts about the electroweak data. Something like a Higgs
boson is required to provide particle masses, but theorists are actively considering alternatives.
The problems of flavour, unification and quantum gravity will require physics
beyond the Standard Model, and astrophysics and cosmology also provide reasons to 
expect physics beyond the Standard Model, in particular to provide the dark matter and explain the
origin of the matter in the Universe. Personally, I find supersymmetry to
be the most attractive option for new
physics at the TeV scale. The LHC should establish the origin of particle masses
has good prospects for discovering dark matter, and might also cast light on unification
and even quantum gravity. Important roles may also be played
by lower-energy experiments, astrophysics and cosmology in the searches for new physics
beyond the Standard Model.
~\\
\begin{center}
CERN-PH-TH/2008-020
\end{center}
\end{abstract}
\end{frontmatter}
%
\section{Status of the Standard Model}
\label{sec:SM}

The Standard Model is in perfect agreement with all confirmed collider data, but
requires a missing ingredient: the Higgs boson or something to replace it in giving
masses to the elementary particles. As well as old theoretical arguments
on high-energy behaviour~\cite{HE},
the consistency of the Standard Model with the precision electroweak data
from LEP, the Tevatron, etc. not only requires something resembling a Higgs boson, but
seems to suggest that it or its replacement physics should be relatively light,
as discussed below.

We learnt in high school that Newton had discovered that  weight is
proportional to mass, and Einstein taught us that energy is related to mass
by the infamous equation $E = m c^2$. However, neither of these
distinguished gentlemen remembered to explain the origin of mass itself.
The answer may be provided by Brout, Englert~\cite{BE} and Higgs~\cite{H}, who postulated a
ubiquitous, universal and constant background field (think a flat desert plain). As Higgs pointed
out, this field would have an associated quantum particle (think grain of sand), 
that we call the Higgs boson,
which has now become the Holy Grail of particle physicists. (Related ideas, applied to
the strong interactions, won half the 2008 Nobel Physics Prize for Nambu~\cite{N}.)

The precision electroweak data by themselves indicate that $m_h = 86^{+36}_{-24}$~GeV~\cite{EWWG},
whereas LEP imposes $m_h > 114.4$~GeV~\cite{LEP}, and the Tevatron collider excludes a Higgs
boson weighing 170~GeV~\cite{Tevatron}. Combining all the available information, the 
Gfitter group finds~\cite{Gfitter}
\begin{equation}
m_h \; = \; 116.4^{+ 18.3}_{- ~1.3} \; {\rm GeV},
\label{Gfitter}
\end{equation}
and quotes the ranges (114, 145)~GeV at the 68\% confidence level and
(113, 168) and (180, 225)~GeV at the 95\% confidence level~\cite{Gfitter}. The Higgs
search is the centrepiece of LHC physics, but will the LHC hare be beaten by the 
Tevatron tortoise? This is possible if $m_h$ is close to the value 170~GeV already
excluded, but seems unlikely if $m_h \sim 115$ to $120$~GeV, which is favoured (\ref{Gfitter}) by
the precision electroweak data and independently by supersymmetry (of which more later).

There is one piece of accelerator data that may disagree with the Standard Model,
namely the measurement of the anomalous magnetic moment of the muon, $g_\mu - 2$~\cite{g-2}.
The Standard Model prediction calculated using older low-energy $e^+ e^-$ data
disagrees with the BNL measurement by over 3 $\sigma$~\cite{LEep}. However, the value
calculated using $\tau$ decay data disagrees with the BNL value by barely 1 $\sigma$,
and the preliminary result of a new analysis of $e^+ e^-$ data from BABAR lies in
between~\cite{BABAR}. Hence it remains unclear whether there is indeed a significant disagreement with
the Standard Model.

\section{Open Questions beyond the Standard Model}

The {\it origin of particle masses}, and whether they are 	due to a Higgs boson, is only one of the
open questions beyond the Standard Model. Another important set of issues bears on the
questions {\it why there are so many different types of matter particles}, and why the quark 
and neutrino flavours mix in the way they do. A related question is {\it the origin of CP violation},
which is described within the quark sector of the Standard Model by the Kobayashi-Maskawa~\cite{KM} 
model, but not really explained. Also, is there some CP violation beyond the Standard Model
that may explain the dominance of matter over antimatter in the Universe today? Another
prominent cosmological question is {\it the nature of the dark matter} that constitutes some 80\%
of the matter in the Universe~\cite{concordance}. 
There is no Standard Model candidate for the dark matter, but
there are good arguments that it might appear at the TeV scale~\cite{DimTeV}. Returning to straight
particle physics, there is the question  {\it whether the fundamental forces may be unified}. If a
unified theory is to include gravity, it must also answer the question  {\it how to construct a
consistent quantum theory of gravity}, a question whose answer has eluded physicists ever
since general relativity and Quantum Mechanics were formulated almost a century ago.

The good news is that many of these questions will be addressed, if not completely
answered, by the LHC. It should discover the Higgs boson, or whatever other physics
replaces it at the TeV scale, which is also the scale at which many dark matter
candidates should appear. If new particles are found at this scale, they will accentuate
the flavour problem and be essential
ingredients that may indicate how to unify the fundamental interactions. Remarkably,
some unification scenarios that invoke additional dimensions of space
predict that gravity should become strong at the TeV scale,
in which case the LHC will be a fantastic place to study quantum effects of gravity.

\section{Supersymmetry}

I make no bones about the fact that I regard supersymmetry as the most plausible
extension of the Standard Model, for many different reasons. It is beautiful, and
apparently an essential ingredient in string theory, the most-favoured candidate
for unifying all the particle interactions including gravity. All well and fine, but there
are, moreover, many reasons for thinking that supersymmetry might appear at the TeV
scale, and hence be accessible to the LHC.

It would help stabilize the hierarchy of mass scales in physics between $m_W$ and
the grand unification scale or the Planck mass, by cancelling the quadratic
divergences in the radiative corrections to the mass-squared of the Higgs boson
\cite{hierarchy},
and by extension to the masses of other Standard Model particles. This
motivation suggests that sparticles weigh less than about 1~TeV, but the exact 
mass scale depends on the amount of fine-tuning that one is prepared to
tolerate~\cite{EOS}. 

Historically, the second motivation for low-scale supersymmetry
was the observation that the lightest supersymmetric particle (LSP) in
models with conserved $R$ parity, being heavy and
naturally neutral and stable, would be an excellent candidate for dark
matter~\cite{Goldberg,EHNOS}. This motivation requires that the lightest
supersymmetric particle should weigh less than about 1~TeV, if it had once
been in thermal equilibrium in the early Universe~\cite{DimTeV}. This would have been the
case for a neutralino $\chi$ (mixture of the supersymmetric
partners of the $Z, \gamma$ and neutral Higgs bosons)
or a sneutrino $\tilde \nu$ LSP, and the argument can
be extended to a gravitino LSP because it may be produced in the decays of
heavier, equilibrated sparticles.

The third reason that emerged for thinking that supersymmetry may be accessible 
to experiment was the observation that including sparticles in the 
renormalization-group equations (RGEs) for the gauge couplings of the Standard Model
would permit them to unify~\cite{GUTs}, 
whereas unification would not occur if only the
Standard Model particles were included in the RGEs. However, this argument does not
constrain the supersymmetric mass scale very precisely: scales up to about
10~TeV or perhaps more could be compatible with grand unification.

The fourth motivation is the fact that the Higgs boson is (presumably)
relatively light, as mentioned earlier. It has been known
for some 20 years that the lightest supersymmetric Higgs boson should weigh
no more than about 140~GeV, at least in simple 
models~\cite{susyH}, 
in perfect agreement with the indications from precision electroweak data and
the unsuccessful searches at LEP and the Tevatron collider.

Fifthly, if the Higgs boson is indeed so light, the present electroweak vacuum would be
destabilized by radiative corrections due to the top quark, unless the Standard Model is 
supplemented by additional scalar particles~\cite{ER}. This would be automatic in supersymmetry,
and one can extend the argument to `prove' that any mechanism to stabilize the electroweak 
vacuum must look very much like supersymmetry.

A sixth argument would be provided by the anomalous magnetic moment of the
muon, $g_\mu - 2$, if we could convince ourselves that the discrepancy between
the data and the Standard Model calculation based on low-energy $e^+ e^-$
data is real. Contributions to $g_\mu - 2$ from supersymmetric particles
would be very capable of explaining any such discrepancy.

\begin{figure}[ht]
\resizebox{0.48\textwidth}{!}{
\includegraphics{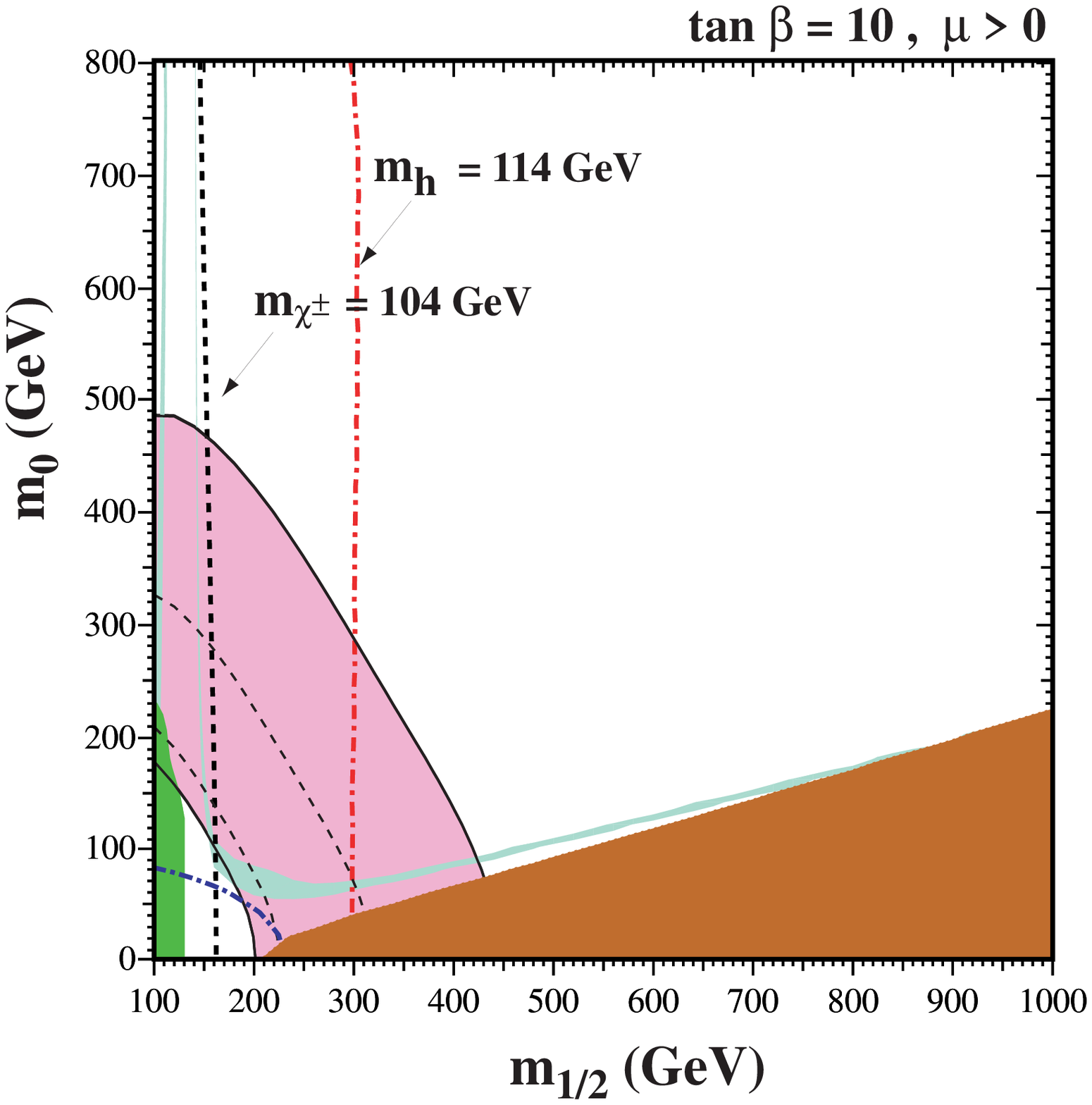}}
\resizebox{0.48\textwidth}{!}{
\includegraphics{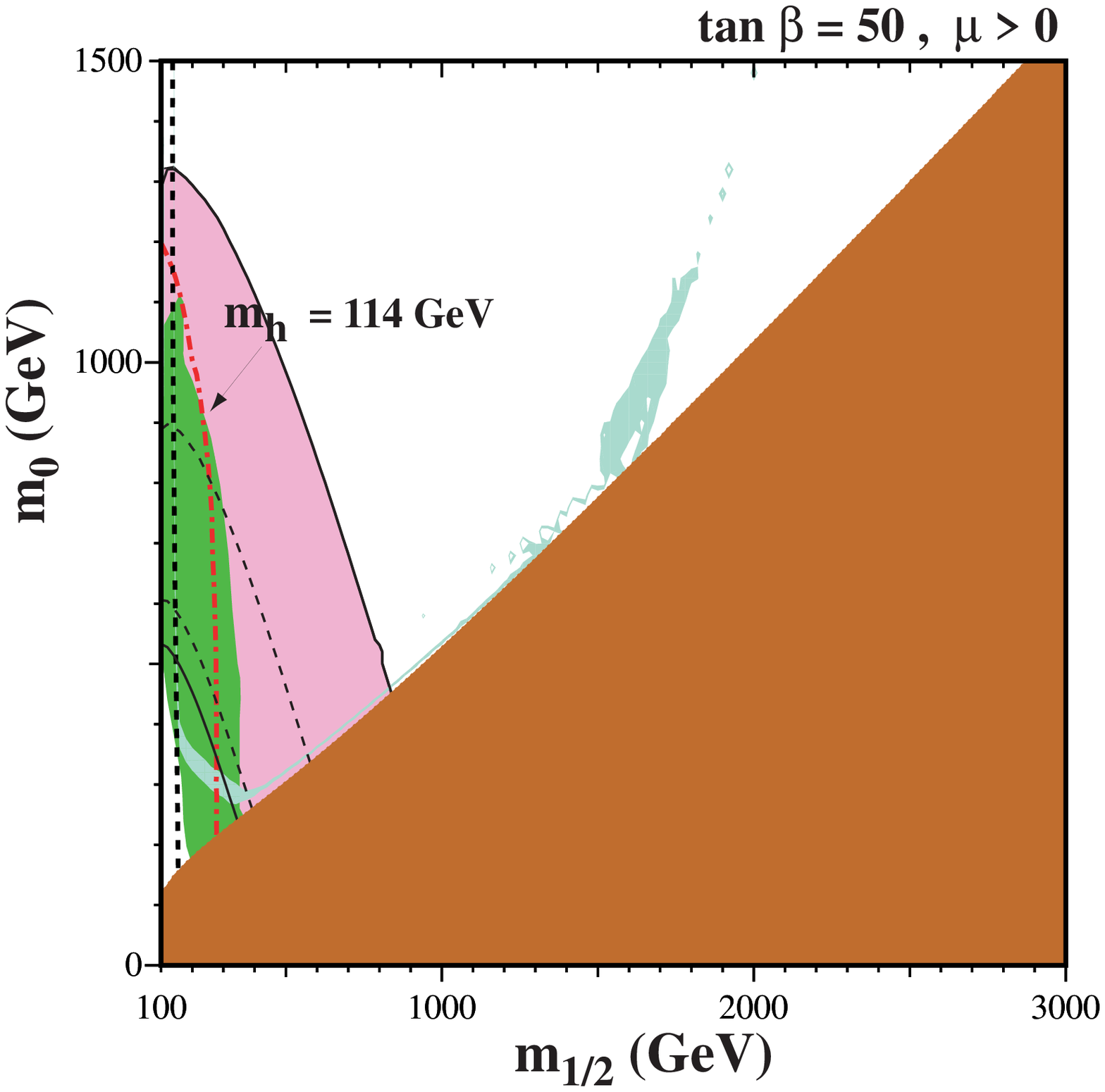}}
\caption{\label{fig:CMSSM1}
{\it The CMSSM $(m_{1/2}, m_0)$ planes for  (a) $\tan \beta = 10$ and (b) $\tan \beta = 50$,
assuming $\mu > 0$, $A_0 = 0$, $m_t = 175$~GeV and
$m_b(m_b)^{\overline {MS}}_{SM} = 4.25$~GeV~\protect\cite{EOSS}. The near-vertical (red)
dot-dashed lines are the contours for $m_h = 114$~GeV, and the near-vertical (black) dashed
line is the contour $m_{\chi^\pm} = 104$~GeV. Also
shown by the dot-dashed curve in the lower left is the region
excluded by the LEP bound $m_{\tilde e} > 99$ GeV. The medium (dark
green) shaded region is excluded by $b \to s
\gamma$, and the light (turquoise) shaded area is the cosmologically
preferred region. In the dark
(brick red) shaded region, the LSP is the charged ${\tilde \tau}_1$. The
region allowed by the measurement of $g_\mu - 2$ at the 2-$\sigma$
level assuming the $e^+ e^-$ calculation of the Standard Model contribution,
is shaded (pink) and bounded by solid black lines, with dashed
lines indicating the 1-$\sigma$ ranges.}}
\end{figure}

Fig.~\ref{fig:CMSSM1} shows the impacts of various constraints on supersymmetry,
assuming that the soft supersymmetry-breaking contributions $m_{1/2}, m_0$ 
to the different scalars and gauginos are each universal at the GUT scale
(the scenario called the CMSSM), and that the
lightest sparticle is the lightest neutralino $\chi$~\cite{EOSS}. We see that narrow strips of
the $(m_{1/2}, m_0)$ planes are compatible with all the constraints, including the 
astrophysical cold dark matter density, and that they vary with $\tan \beta$, the ratio
of supersymmetric Higgs vacuum expectation values.

\section{Where is the Physics Beyond the Standard Model?}

As we have seen, there are many reasons to expect some new physics at the TeV scale,
which should be accessible to the LHC. These include the Higgs boson
and whatever physics (such as supersymmetry) stabilizes the Higgs mass
and hence the electroweak scale. There are also general arguments
that many weakly-interacting particle candidates for dark matter
should also weigh a TeV or less. There is also an abundance of lower-energy
experiments that may be sensitive indirectly to TeV-scale physics.

On the other hand, there are several examples of new physics that may well
lie beyond the LHC's reach. One example is the mechanism responsible for
neutrino masses, which probably lies far beyond the TeV scale, and another is
grand unification, which is thought to occur at a scale $\sim 10^{16}$~GeV. If
one wishes to probe quantum gravity, one may need to reach the
Planck mass $\sim 10^{19}$~GeV, at which gravity in four dimensions would
acquire strength equal to the other particle interactions. Physics at these scales
could only be explored indirectly by the LHC and other low-energy
experiments. On the other hand, grand unification might occur at some
lower energy scale, and gravity might become strong at much lower energies,
if there are additional dimensions of space. There is, to my mind,
no compelling reason to think that extra-dimensional physics should appear
at the LHC, but some possibilities are discussed later in this talk.

Some of the high-scale physics may be accessible only via astrophysics and
cosmology, e.g., via high-energy astrophysical sources, measurements of the
cosmological microwave background radiation, or the detection of
gravitational waves. Thus, although the LHC has bright prospects for
detecting new physics, it does not have a monopoly!

\section{The LHC Haystack}

The total cross section for proton-proton collisions at the LHC is 
$\sim {\cal O}(1/ (100 {\rm MeV})^2$, whereas
cross sections for producing heavy particles weighing $\sim N$~TeV
are typically $\sim {\cal O}(1/ (N~{\rm TeV})^2$. Moreover, many of the cross
sections for producing new particles have 
small coupling factors ${\cal O}(\alpha^2)$, and so are expected to be produced at
rates $\sim 10^{-12}$ of the total cross section or even less. To add insult to injury,
many of the new physics signals have significant backgrounds, and may
require as many as 1000 events to be confirmed. Discovering them at the LHC will be rather like
looking for a needle in $\sim 100000$ haystacks! Hence, big LHC discoveries will
not occur on the first day of high-energy collisions, but will require the
patient accumulation of integrated luminosity and good understanding
of the detectors.

\begin{figure}[htb]
\centering
\epsfig{file=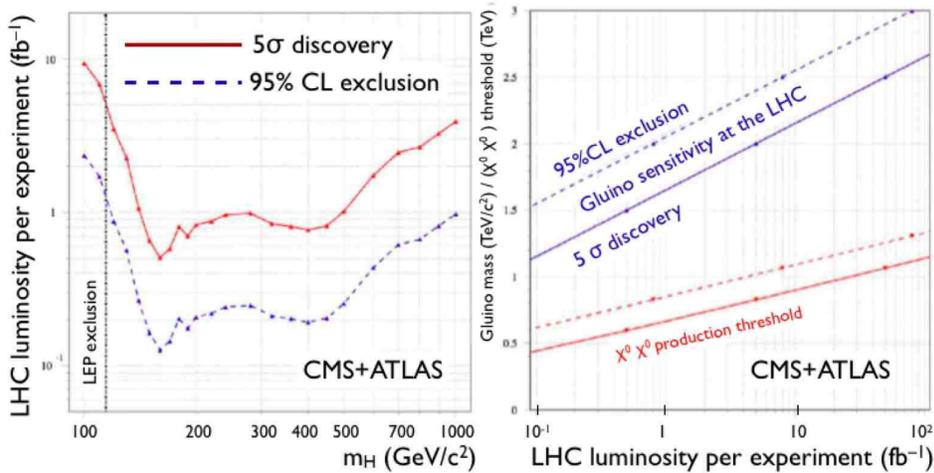,width=5in,angle=0}
\caption{\it The amounts of integrated LHC luminosity at $E_{CM} = 14$~TeV
required (left) either to exclude a Standard Model Higgs boson at the 95\%
confidence level (blue line) or discover it at the 5-$\sigma$ level (red line), 
and (right) either to exclude a gluino (blue dashed line) or discover it (blue
solid line). The corresponding thresholds for $\chi$ pair production in
$e^+ e^-$ are shown in red~\protect\cite{LHCLC}.}
\label{fig:prospects}
\end{figure}

As seen in Fig.~\ref{fig:prospects}, the LHC may start being able to
{\it exclude} certain ranges for the mass of the Higgs boson with a
fraction of an inverse femtobarn of well-understood luminosity. However, the
5-$\sigma$ discovery of a Standard-Model-like Higgs boson,
whatever its mass, will surely require several inverse femtobarns.

\section{Theorists are getting Cold Feet}

The threat that the Higgs boson may soon be discovered, finally, has
concentrated wonderfully the minds of theorists. They are busy concocting
excuses in case it is not discovered, by finding
reasons why the LHC may not discover a Standard-Model-like Higgs boson.

Some are considering composite Higgs models, and grappling with
potential conflicts with the precision electroweak data~\cite{composite}. Others question the
standard interpretation of the electroweak data, and wondering whether
the Higgs boson might be very heavy and hence difficult to detect~\cite{Chanowitz}. Others
accept the electroweak data at face value, but suggest that higher-dimensional 
interactions involving Standard model particles might play an essential
role in the fits, in which case the Higgs boson could again be uncomfortably
heavy~\cite{BS}. Yet others consider Higgsless models, in which $WW$ scattering
becomes strong at high energies~\cite{Higgsless}. This would also create problems with the
precision electroweak data, that may be alleviated by postulating extra
spatial dimensions.

The primary question concerning the precision electroweak data is whether
they all tell the same story, because there are at least two discrepancies
worth noting~\cite{Chanowitz}. One is between the measurements of leptonic and hadronic
asymmetries $A_{L,H}$ in $Z$ decays: relatively low values of $m_h$ are
favoured by $A_L$ (and $m_W$), but larger values would be preferred
by measurements of $A_H$.  Another discrepancy is between the NuTeV
determination of $\sin^2 \theta_W$~\cite{NuTeV} and the value preferred by LEP. Most
of us think that these are statistical or systematic artefacts (the NuTeV
measurement, in particular, is fraught with hadronic uncertainties), and lump
all the data together. On the other hand, perhaps one or the other of these
discrepancies is due to some new physics, in which case one cannot use
the precision electroweak data to estimate $m_h$ within a naive
Standard Model framework~\cite{Chanowitz}.

If, despite these questions, one accepts the electroweak data at face
value, and the Higgs boson is relatively light, are there any alternatives
to the Standard Model or its supersymmetric extension? One interesting
possibility is offered by `Little Higgs' models, in which the Standard-Model-like
Higgs boson is a pseudo-Goldstone boson in some model that reveals
compositeness at the multi-TeV scale~\cite{littleH}. Such models necessarily predict
the existence of additional particles below this compositeness scale,
such as an extra top-like quark, gauge bosons, and Higgs-like scalars.
Lots of exciting stuff for the LHC and other colliders to look for!
Remarkably, one-loop diagrams involving these particles exhibit
cancellations of quadratic divergences, analogous to those in
supersymmetric theories. However, these cancellations do not
continue in higher orders, unlike the case of supersymmetry,
where these divergences occur in all orders of perturbation theory.
This is one of the many reasons why I, personally, prefer
supersymmetry, so let us now cut to the chase.

\section{Searches for Supersymmetry}

As discussed earlier, 
in the minimal supersymmetric extension of the Standard Model (MSSM),
supersymmetric particles carry a multiplicatively-conserved quantum number
$R$, implying that sparticles must be produced in pairs, that heavier sparticles 
must decay into lighter ones, and that the lightest sparticle (LSP) must be stable,
since it has no legal decay mode. The LSP is the supersymmetric candidate
for the astrophysical cold dark matter, and should presumably have neither
electric charge nor strong interactions - otherwise, it would bind to regular
matter and form anomalous heavy nuclei that have never been observed~\cite{EHNOS}.
It is the weakly-interacting nature of the LSP that provides the
classic supersymmetric signature of missing (transverse) energy 
carried away by dark matter particles, e.g., the lightest neutralino $\chi$. 

Simulations for the LHC experiments ATLAS and CMS indicate that this
missing-energy signature should be detectable if the squarks and gluinos
weigh up to about 2.5~TeV~\cite{ATLASCMS}. Assuming that the supersymmetry-breaking
gaugino masses $m_{1/2}$ are universal at the GUT scale, as in the CMSSM, this would
correspond to a mass of about 400~GeV for the neutralino $\chi$. The threshold for
producing pairs of supersymmetric particles must be at least as large as
$2 m_\chi$. Hence discovering (excluding) gluinos at (up to) some mass
would provide a lower limit for producing sparticles in $e^+ e^-$
annihilation, as seen in the right panel of Fig.~\ref{fig:prospects}~\cite{LHCLC}. The
results of LHC searches for supersymmetry will tell $e^+ e^-$ colliders
where (not) to look!

\begin{figure}[ht]
\begin{center}
\resizebox{0.8\textwidth}{!}{
{\begin{rotate}{90}~~~~~~~~~~~~~~~~~~~~~~~~~~~~~~
~~~~~~~~~~~~~~~~~~~{\Large $m_{1/2}$ [GeV]}\end{rotate}}
\includegraphics{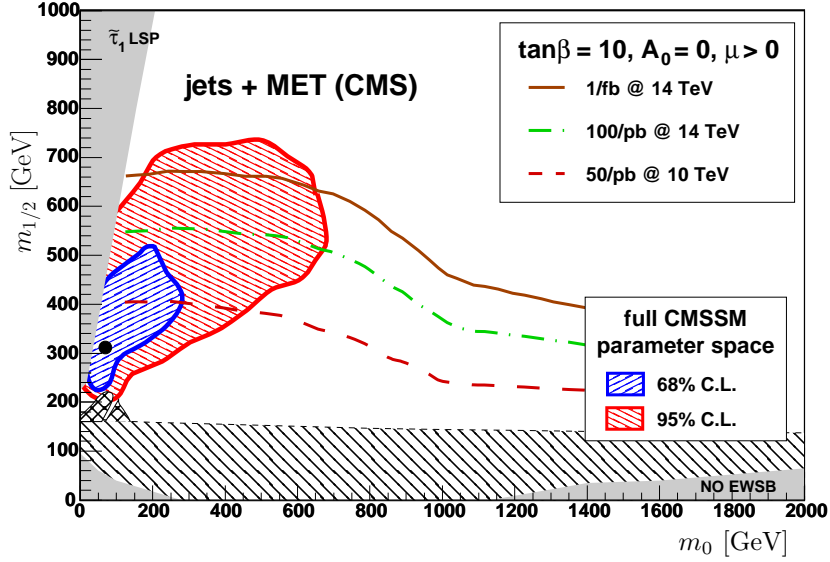}}
\end{center}
\begin{flushright}
{$m_0$~[GeV]}
~~~~~~~~~~~~~~~~~
\end{flushright}
\caption{\label{fig:CMSSM2}
{\it The $(m_0, m_{/2})$ plane for the CMSSM~\protect\cite{MC2},
displaying the best-fit point (black dot), the 68\% C.L. region (blue hatching), the
95\% C.L. region (red hatching) and the region excluded by LEP
and Tevatron searches (black hatching).
Also shown are the 5-$\sigma$ supersymmetry discovery reaches at the LHC assuming 
50/pb of data at 10~TeV, 100/pb at 14~TeV, and 1/fb at 14~TeV, as estimated for the indicated values
of the supersymmetric model parameters.
}}
\end{figure}

How soon might the CMSSM be detected at the LHC? 
Fig.~\ref{fig:CMSSM2} shows the results of a likelihood analysis~\cite{MC2} of the constraints,
crucially including $g_\mu - 2$, assuming that there is a significant discrepancy between
the experimental measurement and the Standard Model prediction as indicated by low-energy
$e^+ e^-$ data~\cite{LEep}.
The black dot indicates the best-fit point in the $(m_{1/2}, m_0)$ plane,
the blue-hatched region is that
favoured at the 68\% C.L., and the red-hatched region is that favoured at the
95\% C.L. Also shown are the regions of the CMSSM $(m_0, m_{1/2})$ plane
where the LHC could discover supersymmetry with the indicated amounts of
luminosity at the indicated energies. We see that the best-fit point would be
accessible already with 50/nb at $E_{CM} = 10$~TeV, and (almost) all the 95\%
C.L. region could be explored with 1/fb at $E_{CM} = 14$~TeV~\cite{MC2}.

\begin{figure}[ht]
\resizebox{0.5\textwidth}{!}{
\includegraphics{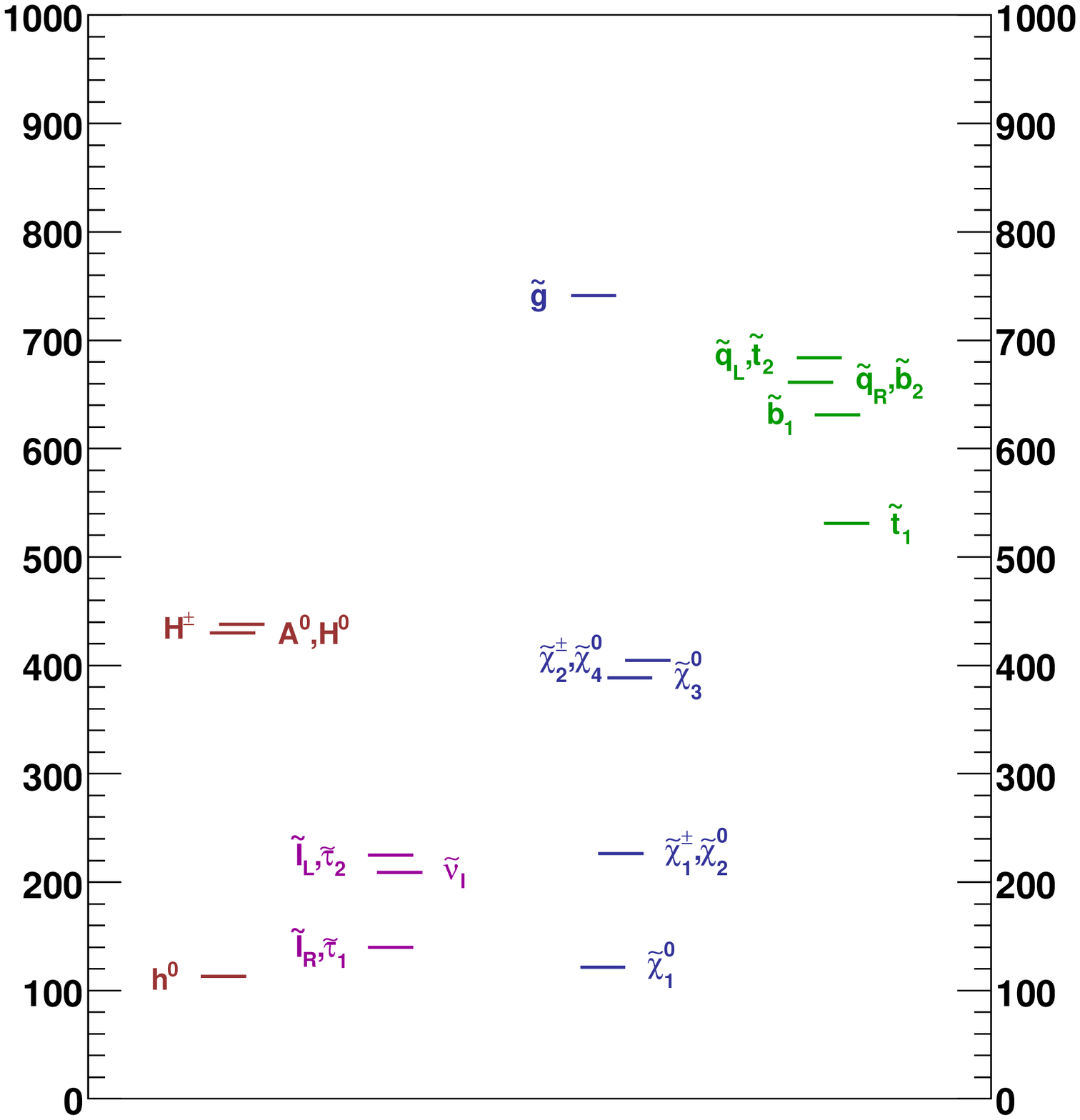}}
\resizebox{0.5\textwidth}{!}{
\vspace{10cm}
\includegraphics{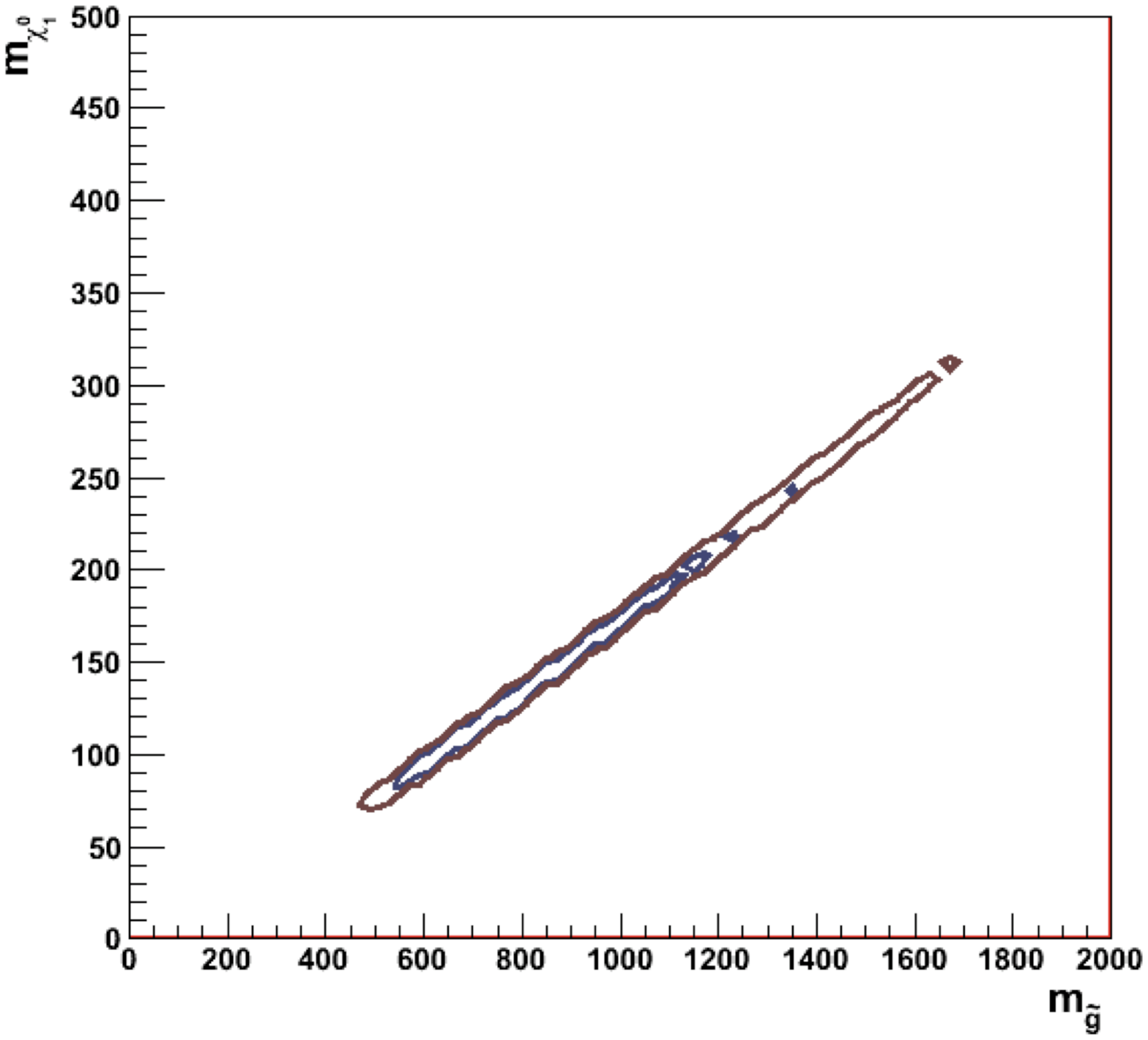}
\vspace{-10cm}}
\caption{\label{fig:spectrum}
{\it Left: The supersymmetric spectrum at the best-fit point in the CMSSM. Right: the correlation
between the masses of the lightest neutralino $\chi$ and the gluino ${\tilde g}$
in the CMSSM~\protect\cite{MC2}.}}
\end{figure}

The left panel of Fig.~\ref{fig:spectrum} displays the CMSSM spectrum at the best-fit point~\cite{MC2}.
We see that some sparticles would be accessible to a linear collider with
$E_{CM} = 500$~GeV, and considerably more with $E_{CM} = 1$~TeV,
and that the full spectrum could be covered with $E_{CM} = 3$~TeV.
The right panel of Fig.~\ref{fig:spectrum} shows the correlation between the mass of
the lightest neutralino $\chi$ and that of the gluino ${\tilde g}$.
They are directly related via the gaugino-mass universality assumption of the CMSSM,
and the other sparticle masses are also highly correlated with $m_\chi$.
The likelihood function is quite asymmetric, being cut off sharply at low
masses, mainly by the LEP lower limit on $m_h$, and more gradually at high
masses, essentially by $g_\mu - 2$. If the $g_\mu - 2$ constraint were weakened,
the likelihood function would rise even more gradually at large masses, and an
infinite mass, i.e., no observable supersymmetry, could not be excluded.

There are various ways to search for astrophysical dark matter, including
searches for antiprotons, antideuterons or positrons in the cosmic rays that are
produced by LSP annihilations in the galactic halo, $\gamma$ rays from
annihilations at the centre of the galaxy, and neutrinos from annihilations
in the centres of the Sum or Earth. There are extant claims to have observed
an excess of positrons and $\gamma$ rays, 
but I do not find them convincing.
There are fewer uncertainties in the direct searches for dark matter scattering
on nuclei deep underground~\cite{direct}, and these might provide the toughest
competition for the LHC in the search for supersymmetry.

\section{CP Violation beyond the Standard Model}

Dirac predicted the existence of antimatter particles with the same masses as
conventional particles, but opposite internal properties such as electric charges.
Antiparticles were duly discovered in cosmic rays and studied using accelerators,
and it came as a complete surprise that matter and antimatter are not quite equal 
and opposite, at least as far as their weak interactions are concerned. This CP
violation can be described within the Standard Model by the mechanism of
Kobayashi and Maskawa~\cite{KM}, who won 
the other half of the 2008 Nobel Physics Prize.
However, some additional matter-antimatter difference would be needed to explain
why the Universe contains mainly matter, not antimatter. Could this be provided
by supersymmetry?

The flavour violation and CP violation seen experimentally are so well
described by the Standard Model, that one normally assumes minimal flavour 
violation (MFV) in the supersymmetric sector, with all squark mixing due to the
Cabibbo-Kobayashi-Maskawa matrix matrix. One therefore assumes that the
soft supersymmetry-breaking scalar masses are universal at the GUT (or
some other high) scale for sparticles with same quantum numbers, but 
universality of the gaugino masses or between scalars having different
internal quantum  numbers is not essential. The MFV model therefore has
the following parameters:
\begin{equation}
M_{1,2,3}; \; \; m^2_{Q, U, D, L, E}, m^2_{H_{1,2}}, \; \; A_{u, d ,e},
\end{equation}
where $Q, U, D, L, E, H_{1,2}$ denote the supermultiplets of the
doublet quarks, singlet quarks, doublet leptons, singlet leptons and Higgses, respectively, and the $A_i$
are soft trilinear supersymmetry-breaking parameters. The maximally 
CP-violating MFV (MCPMFV) model has 19 parameters, of which 6 violate CP,
namely the phases of $M_{1,2,3}$ and of $A_{u, d ,e}$~\cite{MCPMFV}. It is often assumed
that the $Im M_{1,2,3}$ and the $Im A_{u,d,e}$ are universal, but their
non-universality would also be compatible with MFV.

\begin{figure}[ht]
\hspace{ 0.0cm}
\centerline{\epsfig{figure=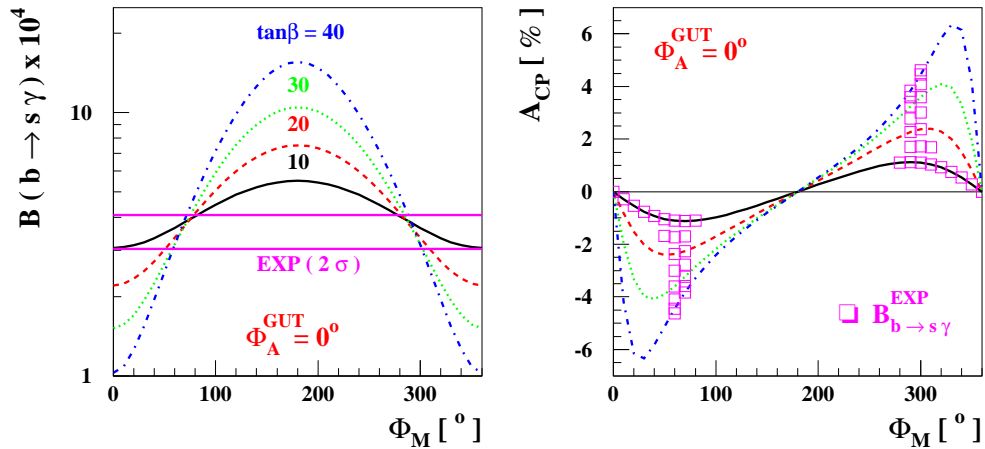,height=14cm,width=14cm}}
\vspace{-6.9cm}
\caption{\it The branching ratio  $B(B \rightarrow X_s \gamma)$ (left)
and the  CP asymmetry $A_{CP}$ (right)  as  functions   of  a common gaugino mass 
phase $\Phi_M$  for  four  values  of
$\tan\beta$, taking the trilinear coupling phase $\Phi_A=0$.   The region allowed
experimentally at the 2-$\sigma$  level is bounded by two horizontal
lines in the  left frame.  In the right  frame, points satisfying this
constraint are denoted  by open squares: see~\protect\cite{MCPMFV} for details. }
\label{fig:MCPMFV}
\end{figure}

The allowed regions of supersymmetric parameter space vary with the
values of the different CP-violating phases in the MCPMFV, which affect, e.g.,
$B_s$ mixing, $B_u \to \tau \nu$ decay and the rate and CP-violating
asymmetry ($A_{CP}$) in $b \to s \gamma$. We have found sample values of the phases
which give values of $A_{CP}$ that are considerably larger than in the Standard
Model, while maintaining consistency with the other $B$-physics observables~\cite{MCPMFV},
as seen in Fig.~\ref{fig:MCPMFV}.

What of the prospects for supersymmetric baryogenesis at the electroweak
scale? The good news is that the electroweak phase transition could be
first order if one of the stop squarks is sufficiently light, and the phases in the
MCPMFV model could in principle provide enough CP violation.
However, both the Higgs and stop masses would be tightly constrained by
the requirement of successful baryogenesis: the available window may soon
be explored by either the Tevatron collider or the LHC~\cite{Carena}.

The parameter space of the MCPMHV model is also tightly constrained
by the experimental upper limits on the Thallium, Mercury and neutron
electric dipole moments (EDMs). However, these three constraints cannot
force all six MCPMFV phases to be small, since there are 
possibilities for non-trivial cancellations between the contributions of different phases~\cite{EDMs}.

\section{Unification and Large Extra Dimensions?}

Unification of all the fundamental interactions was Einstein's dream,
but he never succeeded. One of the ideas he played with was the existence
of extra spatial dimensions, and these play a key role in string theory.
However, as in the case of supersymmetry, string theory does not (yet)
give any clear indications on the scales of any extra dimensions. It used
to be thought that they should necessarily be very small, or order the
Planck length $\sim 10^{-33}$~cm, but this is not necessarily the case. 

How large could extra dimensions be? If they were an inverse TeV in size,
they could break electroweak symmetry (putting the Higgs boson out of
a job) or supersymmetry~\cite{Antoniadis}. If they were micron-size, they would enable the
electroweak hierarchy problem to be rewritten~\cite{ADD}. There are even some 
`warped' scenarios in which there is an extra dimension of infinite size~\cite{RS}:
in this case, we are literally the `scum of the Universe', rather like insects
skating on the surface of a pond, supported by surface tension.

These extra-dimensional scenarios have various possible experimental
signatures at the LHC. Perhaps energy will be lost as it escapes into the extra
dimensions via multiple graviton emission? Perhaps the LHC will
discover Kaluza-Klein excitations of Standard Model particles, whose
wave functions are wrapped around the extra dimensions? Perhaps
gravity will become strong at the TeV scale, in which case some
parton-parton collisions might produce microscopic black holes~\cite{BH}?

These microscopic black holes are generally expected to be very unstable, decaying quickly into
jets, leptons and photons via Hawking radiation. However, there are some 
speculative scenarios in which the black holes might be longer-lived~\cite{CH} or 
even stable: these would be very interesting, but do not panic, because
they would be totally harmless~\cite{LSAG}! The Earth
and other astrophysical bodies have survived being subjected to collisions by
cosmic rays with much higher effective energies than the LHC collisions, and we are
still here.

\section{Conclusions}

There are many good theoretical and astrophysical reasons to
expect physics beyond the Standard Model.
The main raison d'{\^ e}tre of the LHC is to look for such new physics, and it is the tool
of choice for searches for the Higgs boson, supersymmetry and extra dimensions.
However, many other experiments also have key roles to play, e.g., searches
for EDMs, $B$ factories, the Tevatron collider, searches of dark matter, 
high-energy astrophysics and cosmology. Only time will tell who discovers first
which evidence for physics beyond the Standard Model!

%
%
%

%
\end{document}